\documentclass[10pt,a4paper]{article}
\usepackage[T1]{fontenc}
\usepackage{mathptmx}
\usepackage[left=45.35bp,right=45.35bp,top=54bp,bottom=72bp]{geometry}
\usepackage{amsmath,multicol,graphicx,array}
\usepackage[hypcap=false]{caption}
\usepackage[hidelinks]{hyperref}

\raggedcolumns
\makeatletter
\renewcommand{\section}{\@startsection{section}{1}{\z@}%
  {9bp}{4bp}{\normalfont\centering\scshape\normalsize}}
\renewcommand{\subsection}{\@startsection{subsection}{2}{\z@}%
  {5bp}{3bp}{\normalfont\itshape\normalsize}}
\renewcommand{\@seccntformat}[1]{\csname the#1\endcsname.\hspace{0.42em}}
\makeatother

\hypersetup{pdftitle={Selective Verification in Multi-Agent Orchestration with Large Language Models},
  pdfauthor={Ramesh Krishnamaneni; Muralidhar Kurni},
  pdfsubject={Selective verification under a shared evidence failure},
  pdfkeywords={multi-agent systems, large language models, majority voting, selective verification}}
\begin{document}
\vspace*{-48.5bp}
{\centering\fontsize{24}{27}\selectfont Selective Verification in Multi-Agent Orchestration with Large Language Models\par}
\vspace{11bp}
\noindent
\begin{minipage}[t]{.487\textwidth}\centering
\fontsize{11}{13}\selectfont Ramesh Krishnamaneni\\
Principal Solutions Architect\\
\textit{IBM}\\
Dallas, Texas, USA\\
\fontsize{9}{11}\selectfont
\href{mailto:ramesh.krishnamaneni@gmail.com}{ramesh.krishnamaneni@gmail.com}\\
\href{https://orcid.org/0009-0006-7771-1793}{ORCID 0009-0006-7771-1793}
\end{minipage}\hfill
\begin{minipage}[t]{.487\textwidth}\centering
\fontsize{11}{13}\selectfont Muralidhar Kurni\\
Professor of CSE\\
\textit{Anantha Lakshmi Institute of Technology \& Sciences}\\
Ananthapuramu, India\\
\fontsize{9}{11}\selectfont
\href{mailto:muralidhar.kurni@gmail.com}{muralidhar.kurni@gmail.com}\\
\href{https://orcid.org/0000-0002-3324-893X}{ORCID 0000-0002-3324-893X}
\end{minipage}\par

\begin{multicols}{2}

{\small\bfseries\textit{Abstract}---Agents can agree because they relied on the same flawed evidence. We use an exact model of nine binary votes to examine when a verifier should intervene. In the main comparison, workers err 20 percent of the time under ordinary conditions, shared failures occur in 10 percent of cases, and the verifier errs 2 percent of the time in either state. When worker error during a shared failure is 80 percent, checking close votes yields 7.93 percent verdict error, compared with 9.06 percent for random checks at the same expected check rate. The ordering reverses at 90 percent worker error during failure. At the matched check rate for that regime, checking close votes yields 9.98 percent error, random checks yield 9.41 percent, and ranking checks by their expected error reduction yields 6.73 percent. The model also shows that passing branch vote counts preserves the flat majority and reduces the number of messages the root receives. A design guide and a documentation audit of four Python frameworks translate these findings into implementation choices.\par}

{\small\bfseries\textit{Keywords}---multi-agent systems, large language models, selective verification, majority voting, shared evidence failures\par}

\section{Introduction}

Chain-of-thought prompting asks a model to lay out intermediate steps \cite{ref1}. With a team of large language model (LLM) agents, the workflow also determines who acts next, what evidence each worker receives, and when a verdict needs checking \cite{ref2}, \cite{ref3}. These choices can change the verdict and the work needed to reach it.

\subsection{Shared Evidence and Agreement}

Workers can inherit the same retrieval result, premise, or source. If that evidence is wrong, several workers may reproduce the same error. A large vote margin can then reflect a single mistake.

\subsection{Study Design}

We study two decisions that arise when workers submit binary votes. First, how much information must a branch lead pass upstream to preserve the majority? Second, when does checking close votes reduce error more than checking cases at random? We feed the same votes to each aggregation circuit so that changes in the verdict and communication can be traced to the circuit. We then compare verification rules at equal expected check rates.

Exact enumeration isolates the effects of aggregation and verification on synthetic votes. An audit of four Python frameworks identifies documented ways to implement delegation, handoffs, and routing.

\section{Related Work}

Orchestration studies describe how agents obtain and exchange information. Studies of debate and verification examine what that information contributes to the accuracy of a shared verdict.

\subsection{Control and Routing}

Zhu et al. distinguish three control topologies from the choice to make routing adaptive \cite{ref4}. We use that distinction to organize the framework comparison. Emde et al. evaluate frameworks at the system level \cite{ref5}. Kim et al. study how task structure and coordination affect error propagation \cite{ref6}. Sander et al. compare four workflow configurations on a terminal-based benchmark \cite{ref7}. GPTSwarm and Graph-of-Agents adapt communication among agents \cite{ref8}, \cite{ref9}.

\subsection{Debate and Verification}

Debate can improve reasoning, and it can also encourage premature agreement \cite{ref10}, \cite{ref11}. AgentAuditor compares evidence at points where reasoning traces diverge \cite{ref12}. The CAVEAT preprint allocates verifier calls using dependencies among claims and correlated judgments \cite{ref13}. Our analysis holds the votes fixed and derives conditional error rates for each observable margin. This lets us examine when the most valuable checks move from close votes to unanimous ones.

\section{Orchestration Design}

Delegation authority and routing rules are distinct design choices. A supervisor, a hierarchy of leads, and a peer group can each follow fixed transitions or choose the next step from the current run.

\subsection{Authority and State}

Four design choices describe a workflow. The control graph names the permitted delegations and handoffs. The message graph specifies who receives each output. The routing rule uses the current trace and remaining budget to select an action. The state contract says who can modify an artifact, how updates are merged, and where a run can resume.

A supervisor assigns work and combines responses. A hierarchy gives branch leads responsibility for their own workers or subteams. In a peer-to-peer arrangement, several agents can route or hand off work directly. A debate can use peer communication with or without a moderator.

\end{multicols}

\newpage

\begin{multicols}{2}

\subsection{Routing and Communication}

A fixed graph can declare permitted transitions and exit conditions before execution. Conditional routing selects among those transitions using information from the current run. Changing which agents are eligible to receive a task is a further design choice.

Typed messages carry a sender, task identifier, and payload. A shared blackboard gives authorized agents access to common artifacts. Graph-managed state associates fields and merge rules with execution steps and can support checkpoints. For example, a worker can receive a task by message and attach the resulting evidence to a shared record while a state graph tracks completion.

For communication across services, the Agent2Agent (A2A) specification defines how agents discover one another and exchange messages, tasks, and status updates \cite{ref14}. Teams also need to agree on identities, permissions, and the structure of the evidence they exchange.

\end{multicols}

\subsection{Choosing a Control and Verification Policy}

Fig.~\ref{fig1} ties delegation to routing and verification. In its conditional routing example, worker votes go to a gate that can accept the majority or ask a verifier to make the final decision.

\par\noindent\begin{minipage}{\linewidth}
\centering\includegraphics[width=.997\linewidth]{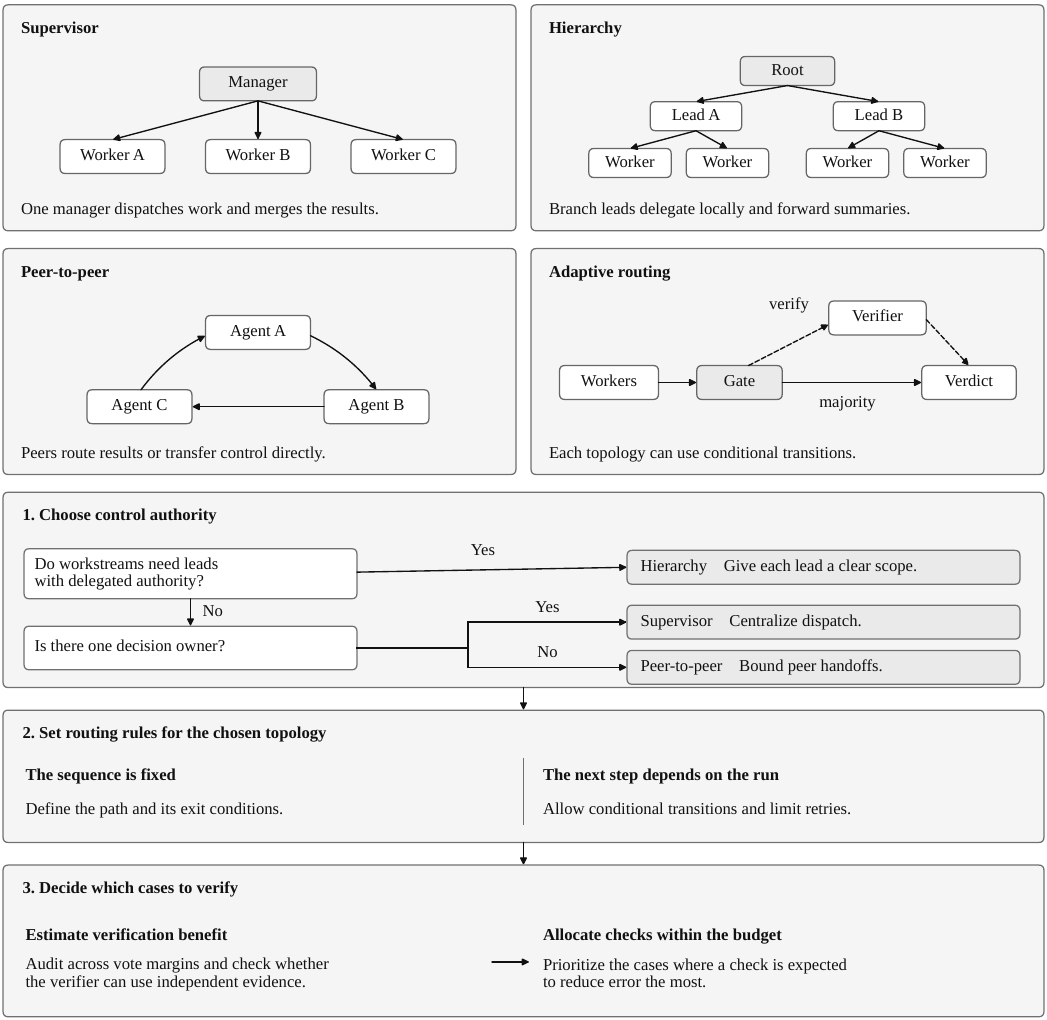}
\captionof{figure}{Control topologies, conditional routing, and selection of verification checks.}\label{fig1}
\end{minipage}\par

\newpage

\begin{multicols}{2}

\section{Framework Capabilities}

The four frameworks expose different controls over who acts next. A team selector chooses a speaker, and a handoff transfers control of the active run. The application still needs a rule for accepting a verdict.

\subsection{Documentation Audit}

We reviewed official Python documentation on November 16, 2026 \cite{ref15}, \cite{ref16}, \cite{ref17}, \cite{ref18}. A named pattern or interface is labeled D. A design assembled from documented features is labeled C. The audit records support at the interface level.

Microsoft has archived AutoGen and directs new projects to Microsoft Agent Framework \cite{ref17}. In existing AutoGen workflows, use TeamTool with serial tool calls. The accompanying evidence matrix records the documentation and available release references.

\end{multicols}

\subsection{Framework Comparison}

Table~\ref{tbl1} maps the four frameworks to the control and routing choices in the design guide. Each cell names a documented feature or a way to compose documented features.

\par\noindent\begin{minipage}{\linewidth}\centering
\captionof{table}{Documented Orchestration Capabilities}\label{tbl1}
\footnotesize\setlength{\tabcolsep}{3.6bp}
\setlength{\arrayrulewidth}{.35bp}\renewcommand{\arraystretch}{1.23}
\begin{tabular}{|>{\raggedright\arraybackslash}p{72.07bp}|>{\raggedright\arraybackslash}p{89.99bp}|>{\raggedright\arraybackslash}p{104.93bp}|>{\raggedright\arraybackslash}p{89.99bp}|>{\raggedright\arraybackslash}p{107.52bp}|}
\hline
\textbf{Framework} & \textbf{Supervisor} & \textbf{Hierarchy} & \textbf{Peer handoff} & \textbf{Adaptive routing} \\ \hline
LangGraph [15] & LangChain subagents as tools\textsuperscript{D} & Nested graphs\textsuperscript{C} & State-driven handoffs\textsuperscript{D} & Conditional edges and Command\textsuperscript{D} \\ \hline
CrewAI [16] & Crew manager\textsuperscript{D} & Leads coordinating crews through a Flow\textsuperscript{C} & Application-level event routing\textsuperscript{C} & Flow router\textsuperscript{D} \\ \hline
AutoGen [17] & SelectorGroupChat for speaker selection\textsuperscript{D} & Nested teams through TeamTool\textsuperscript{C} & Swarm handoffs\textsuperscript{D} & Model-based speaker selection\textsuperscript{D} \\ \hline
OpenAI Agents SDK [18] & Manager calls agents as tools\textsuperscript{D} & Deputies expose worker tools\textsuperscript{C} & Targeted handoffs\textsuperscript{D} & Model- or code-directed routing\textsuperscript{D} \\ \hline
\end{tabular}\end{minipage}\par

\begin{multicols}{2}

\section{Controlled Voting Model}

Correlation among votes changes the usual majority guarantees \cite{ref19}. We represent that dependence with a shared-failure state and reuse the same worker votes across aggregation circuits.

\subsection{Worker Errors and Shared Failure}

The true verdict $Y$ is binary. Each of $n=9$ workers produces a vote $X_i$. A Bernoulli variable $S$ marks a shared failure, with $\Pr(S=1)=q$. Conditional on $S$, workers err independently, with probability $p$ when $S=0$ and $r$ when $S=1$. The model uses the same error rates for either true label. Let $K$ count wrong votes. Then $f_0(k)$ and $f_1(k)$ are the binomial probabilities for $K=k$ in the two states.

Let $U$ count positive votes. The margin $M=\lvert 2U-n\rvert$ equals $\lvert 2K-n\rvert$ for either true verdict. At margin $m$, the possible counts of wrong votes are $k_-=(n-m)/2$ and $k_+=(n+m)/2$. The majority errs only at $k_+$.

\subsection{Aggregation and Response Traffic}

The flat supervisor receives all nine votes. In the hierarchy, each of three leads receives three votes and passes a majority bit or a two-bit count of positive votes. The three branch counts sum to $U$, so the count hierarchy preserves the flat verdict. The majority-bit hierarchy votes on the three branch verdicts. In the peer ring, each vote makes eight hops so that every peer receives all nine votes. The single-worker reference uses $X_1$.

We count deliveries after voting, logical payload bits, and the most deliveries received by one node in a case. Each hop counts once. These measures exclude dispatch and protocol overhead. If $H_j(x)$ is the probability of a wrong majority among $j$ independent votes with error rate $x$, the flat, count, and ring circuits have conditional error $H_9(x)$. The hierarchy with branch majority bits has conditional error $H_3(H_3(x))$. The two states receive weights $1-q$ and $q$.

\subsection{Enumeration Settings}

We enumerate the 512 vote profiles under each state. For a conditional worker error rate $x$, a profile with $k$ wrong votes has weight $x^k(1-x)^{n-k}$. The symmetric grid uses $p\in\{0.10,0.20,0.30\}$, $q\in\{0,0.10,0.25\}$, and $r=1-p$. The stronger failure case uses $p=0.20$, $q=0.10$, and $r=0.90$. The ancillary files contain the enumeration code and exact result tables.

\subsection{Conditional Benefit of a Check}

A check replaces the majority with a verifier's verdict. The verifier errs with probability $\alpha=0.02$ in ordinary cases and with $\beta\in\{0.02,0.50,1.00\}$ during shared failure. Its error is independent of worker votes given $S$. Let $\pi_0=1-q$ and $\pi_1=q$. For state $s$, the joint probability of that state and margin $m$ is $w_s(m)=\pi_s(f_s(k_-)+f_s(k_+))$. With $w(m)=w_0(m)+w_1(m)$, the majority and verifier error rates at margin $m$ are

\begin{equation}
\begin{aligned} e(m)&=\frac{\pi_0 f_0(k_+)+\pi_1 f_1(k_+)}{w(m)},\\ v(m)&=\frac{\alpha w_0(m)+\beta w_1(m)}{w(m)}.\end{aligned}
\end{equation}

Let $\Delta(m)=e(m)-v(m)$ be the expected reduction in error from one check. If $g(m)$ is the probability of checking at margin $m$, the error $R(g)$ and expected checks per case $B(g)$ are

\begin{equation}
\begin{aligned} R(g)&=\sum_m w(m)\bigl(e(m)-g(m)\Delta(m)\bigr),\\ B(g)&=\sum_m w(m)g(m).\end{aligned}
\end{equation}

For equal-cost checks, moving a unit of expected checking from a lower-gain margin to a higher-gain margin reduces error in (2). At a fixed expected budget, the optimal margin-only policy checks in descending order of gain and randomizes within the last margin if needed. When the budget is an upper bound, it can stop once the gain becomes nonpositive.

\subsection{Symmetric Failure and Matched Budgets}

When $r=1-p$, the margin has the same distribution under both states. The two possible counts share a binomial coefficient. Under ordinary conditions, $f_0(k_-)/f_0(k_+)=((1-p)/p)^m$. The resulting majority error rate is

\begin{equation}
e(m)=q+\frac{1-2q}{1+\left((1-p)/p\right)^m}.
\end{equation}

For $0<p<1/2$ and $q<1/2$, (3) falls as $m$ grows, while verifier error stays at $(1-q)\alpha+q\beta$. Near ties then offer the largest expected gain. At each cutoff $t\in\{0,1,3,5,7,9\}$, the near-tie rule checks cases with $M\leq t$. The random rule checks each case with probability $\Pr(M\leq t)$. The rule ranked by expected gain uses the model probabilities and the same expected budget.

\section{Results}

The aggregation circuit determines how vote traffic is distributed. The verification rule determines which verdicts get a second decision for a given expected number of checks.

\subsection{Response Traffic and Aggregation}

Table~\ref{tbl2} compares verdict error and response traffic for the five circuits at $p=0.20$ and $r=0.80$. Forwarding counts reduces the deliveries received by the root from nine to three and preserves the flat majority on every profile. The hierarchy makes twelve deliveries in total.

\par\noindent\begin{minipage}{\linewidth}\centering
\captionof{table}{Routing Results with Verdict Error Reported in Percent}\label{tbl2}
\footnotesize\setlength{\tabcolsep}{2.1bp}
\setlength{\arrayrulewidth}{.35bp}\renewcommand{\arraystretch}{1.23}
\begin{tabular}{|>{\raggedright\arraybackslash}p{73.08bp}|>{\centering\arraybackslash}p{26.27bp}|>{\centering\arraybackslash}p{30.26bp}|>{\centering\arraybackslash}p{22.29bp}|>{\centering\arraybackslash}p{28.26bp}|>{\centering\arraybackslash}p{34.54bp}|}
\hline
\textbf{Circuit} & \textbf{q = 0\newline (\%)} & \textbf{q = 0.10\newline (\%)} & \textbf{Hops} & \textbf{Bits} & \textbf{Max.\newline received} \\ \hline
Single worker & 20.00 & 26.00 & 0 & 0 & 0 \\ \hline
Flat supervisor & 1.96 & 11.57 & 9 & 9 & 9 \\ \hline
Hierarchy (majorities) & 3.02 & 12.42 & 12 & 12 & 3 \\ \hline
Hierarchy (counts) & 1.96 & 11.57 & 12 & 15 & 3 \\ \hline
Peer ring & 1.96 & 11.57 & 72 & 72 & 8 \\ \hline
\end{tabular}\end{minipage}\par

Forwarding branch majority bits changes the verdict on 54 of the 512 profiles. With branch counts of $(2,2,0)$, the root sees two positive branch verdicts while the flat majority sees four positive worker votes. The ring's 72 deliveries reflect forwarding every vote to every peer.

\subsection{Sensitivity to the Severity of Shared Failure}

Table~\ref{tbl3} compares verification policies at the check rate induced by $M\leq 3$ in each regime, with $p=0.20$, $q=0.10$, and $\alpha=\beta=0.02$. The last column is majority error conditional on unanimity before verification. Unchecked majority error across all votes is 11.57 percent at $r=0.80$ and 11.75 percent at $r=0.90$.

\par\noindent\begin{minipage}{\linewidth}\centering
\captionof{table}{Verification Outcomes at the Near-Tie Budget}\label{tbl3}
\footnotesize\setlength{\tabcolsep}{2.1bp}
\setlength{\arrayrulewidth}{.35bp}\renewcommand{\arraystretch}{1.23}
\begin{tabular}{|>{\raggedright\arraybackslash}p{44.20bp}|>{\centering\arraybackslash}p{32.25bp}|>{\centering\arraybackslash}p{31.25bp}|>{\centering\arraybackslash}p{31.25bp}|>{\centering\arraybackslash}p{35.24bp}|>{\centering\arraybackslash}p{40.51bp}|}
\hline
\textbf{Worker error\newline in failure} & \textbf{Cases\newline checked (\%)} & \textbf{Near-tie\newline error (\%)} & \textbf{Random\newline error (\%)} & \textbf{Ranked\newline error (\%)} & \textbf{Unanimous\newline error (\%)} \\ \hline
80\%\newline Symmetric & 26.15 & 7.93 & 9.06 & 7.93 & 10.00 \\ \hline
90\%\newline Stronger & 24.06 & 9.98 & 9.41 & 6.73 & 24.28 \\ \hline
\end{tabular}\end{minipage}\par

Verifier quality also matters. With $r=0.80$, $q=0.10$, and $\beta=1$, checking every case gives 11.80 percent error. The lowest error among the evaluated cutoffs is 10.39 percent at $M\leq 1$, which checks 8.26 percent of cases. Without shared failures, the same cutoff with the 0.02-error verifier reduces error from 1.96 to 0.47 percent.

\end{multicols}

Fig.~\ref{fig2} shows how this comparison changes with the expected share of cases checked. The gain-ranked and near-tie curves coincide in the symmetric model. In the stronger failure case, the gain-ranked rule checks unanimous cases first.

\par\noindent\begin{minipage}{\linewidth}
\centering\includegraphics[width=.997\linewidth]{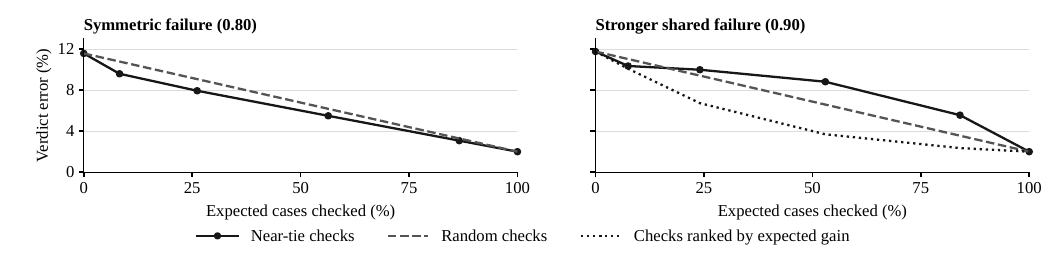}
\captionof{figure}{Verdict error at six expected check rates. Lines connect the evaluated rates. Ordinary worker error is 0.20, the shared-failure probability is 0.10, and verifier error is 0.02 in both states.}\label{fig2}
\end{minipage}\par

\newpage

\begin{multicols}{2}

\section{Discussion}

Vote counts preserve a binary majority exactly. Using the vote margin to allocate checks also requires evidence about the error rates of the majority and the verifier.

\subsection{What to Measure in a Live System}

A workflow that needs to review supporting evidence should preserve where that evidence came from. Tracking the sources shared across branches can help explain why several workers made the same mistake.

With trusted labels, teams can sample cases across vote margins and estimate how often the majority and verifier err. A held-out set can test whether those estimates improve check allocation. A live comparison should hold tasks, tools, and model access fixed and include a single-agent workflow at the same resource budget. Record verdict accuracy, token use, and latency, including failed attempts and retries.

\subsection{Model Limits}

The model assumes binary votes, equal-cost checks, conditionally independent worker errors, and deterministic delivery. The severity comparison holds the frequency of shared failure fixed. Dependencies within a state, differences in worker skill, and variable check costs would change the allocation problem. The framework audit reports documented capabilities. Token use, latency, and accuracy in deployed frameworks remain empirical questions.

\section{Conclusion}

The model identifies the information needed to reproduce a binary majority and the conditions under which a close vote is worth checking first.

\subsection{Preserving the Majority}

A branch can forward its positive vote count and leave the majority decision to the root. In the three-branch example, this reduces the deliveries received by the root while reproducing the flat verdict on every vote profile.

\subsection{When to Spend a Check}

The value of a check depends on how the vote margin relates to shared failure and verifier error. Close votes are valuable targets in the symmetric model. When shared evidence is sufficiently misleading, the largest benefit can come from checking unanimous decisions.

\section*{Acknowledgment}

Text-to-text generative AI assisted with manuscript drafting, source discovery, and development and checking of the synthetic benchmark. The enumeration code generated the reported numerical results.

\begingroup\footnotesize\setlength{\parskip}{0pt}

\endgroup

\end{multicols}

\end{document}